\documentstyle[aas2pp4]{article}

\input{psfig}

\begin{document}
\bigskip

%
\title{
On Rapidly Rotating Magnetic Core-Collapse Supernovae}
 
\author{J. R. Wilson$^{1,2}$,    G. J. Mathews$^{2}$ and H. E. Dalhed$^{1}$}
\begin{center}
$^{1}$University of California,          \\
Lawrence Livermore National Laboratory\\
Livermore, CA, 94550                  \\
$^{2}$Center for Astrophysics, University of Notre Dame, \\
Notre Dame, IN 46556, U.S.A.
\end{center}

\date{\today}
\begin{abstract}
 We  have analyzed the magnetic effects that may occur in rapidly rotating 
core collapse supernovae.
We consider effects from both magnetic turbulence and the formation
of magnetic bubbles.
For magnetic turbulence we 
have made a perturbative analysis for our spherically symmetric
 core-collapse supernova model that incorporates the
build up of magnetic field energy in the matter accreting onto
the proto-neutron star shortly after collapse and bounce.
This significantly modifies the pressure profile and
increases the heating of the material above the proto-neutron star 
resulting in an explosion even in rotating stars that  would not explode otherwise.
Regarding magnetic bubbles we show  that a 
model with an initial  uniform magnetic
field  ($\sim 10^{8}$)   G    and uniform
angular velocity of $\sim  0.1$  rad   s$^{-1}$
can form magnetic bubbles 
due to the very non homologous nature of the collapse.
It is estimated that the buoyancy of the bubbles
causes matter in the proto-neutron star to rise, carrying 
neutrino-rich material to the neutron-star surface.  
This increases the neutrino luminosity sufficiently at early times 
to achieve a successful neutrino-driven explosion. Both magnetic mechanisms
thus provide new means for initiating
a Type II core-collapse supernova.
\end{abstract}

\keywords{Gravitation - Hydrodynamics - Instabilities - Stars: rotation - Supernovae: general}

\section{Introduction}

     In most supernova models with  pure spherical symmetry, 
    after  a massive star collapses due to the exhaustion of its nuclear fuel,
the neutrino luminosity from the
     proto-neutron star (PNS) is too low to heat the in-falling material sufficiently
     to expel matter from the star (e.g.~Bruenn 1985; 1993,  
Burrows, Hayes, \& Fryxell 1995;  Yamada et al. 1999;
Fryer \& Heger 2000; Rampp \& Janka 2000, 2002; 
Liebend\"orfer et al. 2001;
Mezzacappa et al 2001;
Akiyama et al. 2003
Buras et al. 2003; 
Thompson, Burrows, \& Pinto 2003;
Burrows 2004; Cardall 2004).
However,  the  Livermore supernova model 
(cf.~Wilson \& Mayle 1988; 1993; Wilson \& Mathews 2003)
avoids this problem and 
is able to explode in spherical symmetry by inducing
a larger amount of neutrino heating soon after the core bounce.  
  For this reason it is important to examine any possible
  means to induce additional heating above the proto-neutron star.
  
One important mechanism  
for such heating, for example,  is that   
 the proto-neutron star can become hydrodynamically unstable
     a few hundred milliseconds after the core bounce
 due to the so-called neutron-finger instability (Wilson \& Mayle 1988; 
Wilson \& Mathews 2003). 
This instability results from the build up of dense material with a large
neutron-to-proton ratio near the surface of the proto-neutron
     star.  Sufficiently neutron-rich material can overcome the buoyancy 
caused by the high entropy near the surface.  As surface material sinks downward
neutrino-rich material rises to the surface.  This enhances the neutrino luminosity
and produces enough heating of material behind the
shock to produce an explosion (cf.~Wilson \& Mathews 2003).

The fact that other models 
of Type II supernovae do not 
     exhibit this instability can be attributed
to a number of possibilities (e.g.~Bruenn et al. 2004) 
such as differences in the equation of state
 employed, the detailed way in which convection is treated,
and/or  the treatment of neutrino flow. Hence,
this mechanism is  controversial (Bruenn et al.~2004) as a means to induce core-collapse
supernovae.  
THerefore, in this paper  we describe first schematic
calculations of some plausible, and perhaps more compelling,
alternatives to the neutron-finger instability
to overcome the lack of sufficient neutrino luminosity at early times in the explosion.
We have investigated two  magneto-hydrodynamic (MHD)
     effects both above and below  the rotating proto-star surface
 that may be strong enough to enhance the
     neutrino luminosity and produce an explosion. These  processes
     could either increase the neutron-finger instability or replace it in
     models that have no surface convection.

\section{ Models}

  The purpose of the present paper is to make a schematic study of
the possible roles of magnetohydrodynamics in rotating core-collapse
supernovae.  While the models we utilize are adequate to illustrate 
the order of magnitude of these 
effects, we emphasize that it will be necessary  to do this calculation
in two or three spatial dimensional 
magneto-hydrodynamics (MHD) to prove that the ideas presented
here are correct.  
Such calculations, however,  present an exceedingly
difficult computational challenge.
 For the purposes of the present schematic investigation, 
however, a much simpler model is employed which involves the average 
spherical effects of the inherently multidimensional rotation
and magnetic effects as a perturbation on
one-dimensional hydrodynamics.  This is a reasonable
approximation as a means of exploring the parameter space and
extracting the essential physics
as long as we are considering moderate rotation rates and
relatively weak but realistic magnetic fields.

As we shall see, the main effects in our
perturbation analysis  are 
build up of magnetic turbulence and field energy above the proto-neutron star 
shortly after collapse and bounce due to the accretion of magnetized matter,
and the formation of magnetic bubbles and magnetic-driven convection below the surface
of the PNS.  We demonstrate that both of these effects can
 significantly impact the explosion mechanism.

\subsection{Supernova Model}
Details of the current version of the Livermore supernova model have been described
in Wilson \& Mathews (2003).  For completeness we here summarize
the basic physics and the  way in which the effects
of rotation and MHD are implemented.
To begin with the metric for a spherical neutron star
is written in Lagrangian coordinates,
\begin{eqnarray}
ds^2 &=& -a^2\biggl[1 - \biggl({U \over \Gamma}\biggr)^2\biggr] dt^2 
- {2 a U \over \Gamma^2} dr dt  + {d r^2 \over \Gamma^2}
\nonumber \\
&+& r^2(d \theta^2 + \sin^2{\theta} d \phi^2)~~,
\end{eqnarray}
where $a$ is the inverse of the time component of four velocity
 $a \equiv  1/U^t$.  It is, thus, related to the 
gravitational red shift. 
The quantity $r$ is a distance coordinate with
proper distance is given by
\begin{equation} 
{\rm Proper~Distance} = \int {dr \over \Gamma}~~.
\end{equation}
where,
\begin{equation}
\Gamma \equiv  \biggl( 1 + U^2 - {2 M \over r} \biggr)^{1/2}~~.
\label{newgamma}
\end{equation}
The quantity  $M$ is the gravitational mass interior to
$r$ as defined below, and
$U^2  = U^rU_r$ is the square of the  radial component of the four velocity.

For the metric coefficient $a$ 
the vanishing of the covariant derivative
 $T^{j \mu}_{~~;\mu} = 0$ implies,
\begin{eqnarray}
a &=& exp \biggl[ \int^{m_{max}}_m {dm \over \rho h}
\biggl( {\partial P_{eff} \over \partial m}  +
{b r^2 \rho \over a} {\partial \over \partial t} 
\biggl\{ {\Phi_\nu \over r^2 \rho^2}\biggr\}
\nonumber \\
&-& {2 b  \Gamma \over r} W_\nu \biggr) \biggr] ~~,
\label{metrica}
\end{eqnarray}
where $m_{max}$ is the mass coordinate at the boundary of
the numerical grid.  The quantity
 $h \equiv 1 + \epsilon + P/\rho$ is the relativistic enthalpy, and
\begin{equation}
b \equiv {1 \over 4 \pi r^2 \rho}~~.
\end{equation}
The quantities $\Phi_\nu$ and $W_\nu$ refer to the angle integrated neutrino
flux and a nonthermal neutrino pressure correction factor, respectively,
as described in Wilson \& Mathews (2003).

\subsection{Matter Equations}
For the present application we write the radial four acceleration as,  
\begin{eqnarray}
{1 \over a} {\partial U \over \partial t} & = & 
-{ \Gamma \over \rho h}
\biggl[ {1 \over b} {\partial P_{eff}
 \over \partial m}  +
{ r^2 \rho^2 \over a} {\partial \over \partial t}
\biggl\{ {\Phi_\nu \over r^2 \rho^2}\biggr\}
\nonumber \\
&& 
- {2   \Gamma \over r} W_\nu \biggr] 
- {M \over r^2} -  4 \pi r P_{eff} ~~.
\label{udot}
\end{eqnarray}

As discussed below, effects of rotation and
magnetic field energy are absorbed into an effective
pressure in the acceleration equation, i.e. we write:
\begin{equation}
P_{eff} = P_M + P_\nu + P_{rot} + P_{mag}~~,
\label{peff}
\end{equation}
where $P_M$ and $P_{\nu}$ are the usual contributions
from matter in thermal equilibrium and neutrinos
which are nonthermal and must be transported
explicitly.  The effective pressure perturbations 
from rotation and magnetic field energy, $P_{rot}$ and $P_{mag}$, are 
defined below  in section \ref{magturb}.

 The condition of baryon number conservation
leads to
auxiliary equations for the matter evolution:
\begin{equation}
\rho = { 1 \over b} {\partial r \over \partial m}~~.
\end{equation}
\begin{equation}
{1 \over a} {\partial \rho \over \partial t} = -\rho {1 \over r^2}
 {\partial  \over \partial r} (r^2 U) + {1 \over 2 \Gamma} \rho r \Phi_\nu~~.
\label{rhodot}
\end{equation}
The gravitational mass is given by
\begin{equation}
M = 4 \pi \int_0^m dm {\partial r \over \partial m} r^2 
\biggl[ \rho (1 + \epsilon) + {U \over \Gamma} \Phi_\nu \biggr]~~,
\end{equation}
with 
\begin{equation}
{1 \over a} {\partial M \over \partial t} = 4 \pi  r^2
(U P + \Gamma \Phi_\nu)~~,
\end{equation}
and
\begin{equation}
{1 \over b} {\partial M \over \partial m} = 4 \pi  r^2
\biggl[ \Gamma \rho (1 + \epsilon) + U \Phi_\nu\biggr]~~.
\end{equation}
The baryon rest mass of the star is then simply  given by the
integral over the proper volume, $d(Vol) = 4 \pi r^2 dr/\Gamma$,
\begin{equation}
M_0 = 4 \pi \int r^2 dr {\rho \over \Gamma}~~.
\end{equation}

The matter internal energy evolves according to
\begin{equation}
{1 \over a} {\partial \epsilon_M \over \partial t}
 = -P_M {1 \over a} {\partial \over \partial t} \biggl( 
{1 \over \rho}\biggr) - {1 \over \rho}
\sum\limits_{i=1}^6 \int \Lambda_i dE d\Omega_\nu~~,
\label{edot}
\end{equation}
where $P_M$ is the matter pressure and the $\Lambda_i$ are various 
neutrino scattering and absorption source
terms (Wilson \& Mathews 2003). The neutrino transport
is treated with
appropriate flux-limited diffusion.

The condition of lepton number conservation
leads to an expression for the change in
the average electron fraction (or charge per baryon) $Y_e$
due to weak interactions,
\begin{equation}
{\rho \over m_b} {1 \over a} {\partial Y_e \over \partial t} = 
- \sum_i (\Lambda_i - \bar \Lambda_i) {dq \over q} d\Omega_\nu~~,
\label{lepton}
\end{equation}
where $q \equiv a \epsilon_\nu$, $\epsilon_\nu$ is the neutrino energy,
and $q$ is the energy a neutrino would have if it was removed to infinity.

\subsection{Magnetic Fields}
Magnetic fields are easily added to the simulation via the
electromagnetic stress-energy tensor,
\begin{equation}
T_{\mu \nu} = T_{\mu \nu}^{\rm Fluid} + T_{\mu\nu}^{EM}~~,
\end{equation}
where,
\begin{equation}
T_{\mu\nu}^{EM} = {1 \over 4 \pi} (g_{\alpha \mu} F^{\alpha \beta} F_{\beta \nu  }
 - {g_{\mu \nu } \over 4} F^{\alpha \beta } F_{\alpha \beta})
~~,
\end{equation}
and as usual, the electromagnetic tensor
$F_{\mu \nu}$ can be related to a vector potential $A_\nu$,
\begin{equation}
F_{\mu\nu}={\partial A_\nu \over \partial x^\mu} -
{\partial A_\mu \over \partial x^\nu}~~.
\end{equation}
In cylindrical symmetry the non-vanishing spatial components of $F_{\mu \nu}$ are thus,
\begin{equation}
F_{r z}=H_\phi~~~~,~~~~ F_{r\phi}={\partial A_\phi \over \partial r}~~~~,~~~~
F_{z \phi}={\partial A_\phi \over \partial z}~~.
\end{equation}
Then, from the  assumption of
perfect conductivity,
$U^\mu F_{\mu\nu}=0$, the space-time components can be obtained.
\begin{equation}
F_{tr} = V^z H_\phi + V^\phi F_{r\phi}
\end{equation}
\begin{equation}
F_{tz} = V^\phi F_{z\phi} - V^r H_\phi
\end{equation}
\begin{equation}
F_{t\phi}=-V^r F_{r\phi} - V^z F_{z \phi} = {\partial A_\phi \over \partial t}
\end{equation}

The time evolution of $H_\phi$  then can be deduced from
Maxwell's equation
\begin{equation}
 F_{rz ;t} + F_{tr ;z} + F_{z t ;r}  = 0~~,
\end{equation}
 which gives,
\begin{equation}
{\partial H_\phi \over \partial t} =
{\partial \over \partial z} \biggl(V^\phi {\partial A^\phi \over \partial r}\biggr)
 - V^z H_\phi -
{\partial \over \partial r} \biggl(V^\phi {\partial A^\phi \over \partial z}\biggr)
+ V^r H_\phi~~.
\end{equation}
In the present work we ignore the back reaction of the magnetic fields on the matter fluid
in the simulations, but estimate their effects below in Section 3.
 
\subsection{ Initial Conditions}
     To evolve the matter equations of motion \ref{udot}-\ref{lepton},
 we adopt 250 nonuniform radial zones.  The grid extends to $\sim 170$ 
zones above the photosphere.
For the magnetic evolution we
adopt 30 angular zones in a 90$^o$ quadrant so that the size of each angular zone 
is 3$^o$.  In our calculations the region of magnetic-driven turbulence extends over a region
of $\sim 10-20$ radial zones.  This resolution has been employed and tested in previous 
supernova collapse and rotating star calculations (Wilson \& Mathews 2003) and should be
adequate for the analysis here.
  
  The  initial MHD model assumes that the star
is rotating with a uniform angular velocity in
     the inner 5 M$_\odot$ of the star.  It is also
 threaded by a uniform
     magnetic field in the direction of the axis of rotation
 before the start of the dynamic core-collapse phase. 
  Various initial models have been explored.  For 
most of the models reported on here,
the strength of the initial  magnetic field is chosen such that the final neutron star 
would have a surface magnetic field of $H \sim 10^{12}$  G. 

The initial rotational velocity  in some models leads  to a post collapse
rotation period ($P \sim 1.4$ ms). 
This is close to the shortest possible  Keplerian neutron-star rotation period
     ($P ^<_\sim 1$ ms, Burgio, Schulze \& Weber 2003), and is
comparable to the minimum values 
in the observed period distribution for pulsars  (Phinney \& Kulkarni 1994; Weber 1999;
Manchester 2004).  Observed young pulsars (Manchester 2004)
 in supernova remnants have much longer periods 
($\sim 1$   s) than that
obtained in these calculations. However they also have large spin down rates. 
     Indeed, the stars modeled here should also
have a large spin-down rate.  Since the poloidal field threads the rapidly rotating neutron star,
     and we use a massive outer envelope of very low angular velocity, a very
     high torque should develop between the spinning neutron star and the
outer star plus magnetic field.  These torques are sufficient to
slow the neutron star down to  well within the observed range by the time it becomes 
an observable pulsar.

\subsection{Magnetic Turbulence above the PNS}
\label{magturb}
Above the proto-neutron star for a few tenths of a second after bounce 
is a region below
the bounce shock front and above the almost static proto-neutron 
star radius where matter is slowly accreting.
At post bounce times typically $t_{pb} \sim 200$ ms, the proto-neutron star radius is $\sim 40 $ km and the shock radius is $\sim 170$ km.
Later,  at $t_{pb} \approx 300$ ms, the proto-neutron star radius is 32 km
and the shock radius has contracted to 140 km.  The Mach number
in the sub-shock region is $<0.1$.

In the sub-shock region magnetic field generation is possible for a 
rapidly rotating magnetized collapsing star.  To model the
evolution of the magnetic field, we follow the general principles given in Balbus and Hawley (1998) [hereafter BH98].

\begin{figure}
\psfig{file=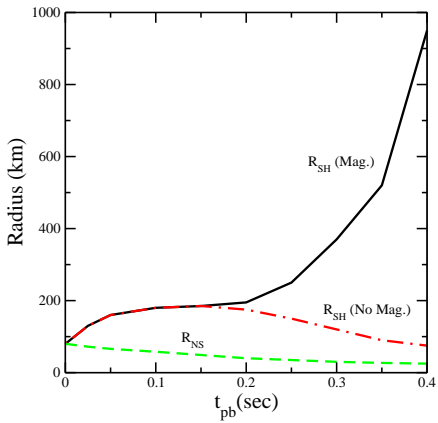,width=3.5in,angle=270}
\caption[]{ The lowest curve shows the radius of the proto-neutron star
as a function of time post bounce.  The upper curves give
the radius of the  accretion shock for models with
an initial angular velocity of $\omega = 0.071$ s$^{-1}$
with and without magnetic turbulence as labeled.
The upper curve clearly shows the effects of magnetic field
amplification on the shock front.  The lower curve shows that rotation alone has little
effect for this particular model.}
\label{fig:1}
\end{figure}

The work of BH98  
has led us to examine the stability of the accretion flow
of matter in the waist region shortly after core bounce.  After bounce
a shock moves out above the proto-neutron star.  This produces
a region of slowly moving (Mach number $< 0.1$) matter 
(see Figure \ref{fig:1}).  For an initially uniformly
rotating iron core this subshock region has an angular 
velocity profile of $\omega \propto r^{-1.8}$.
While this velocity profile is different from the Keplerian profile
($\omega \propto r^{-3/2}$) studied
in BH98, it is still unstable to magnetohydrodynamic flow.
BH98 give a maximum growth rate of $\lambda = (r/2) (d\omega/dr)
\approx 0.9 \omega$.  We adopt this growth rate.
Hence we write: $\dot H = \lambda H$,

BH98 treat accretion disks that are almost
static and orbital velocities that are close to Keplerian.
However, in our supernova model the accretion is 
rapid and the angular velocity is not Keplerian.  

We assume that the turbulent magnetic field amplitude then grows as
\begin{equation}
\tilde H(t) = H_0(r) e^I~~,
\end{equation}
where $I$ is the integrated growth rate
\begin{equation}
I = \int \lambda dt = 0.9 \int \omega dt = 0.9 \int_{r}^{r_{sh}} \biggl|\frac{\omega}{v} \biggr|dr~~,
\end{equation}
where $r_{sh}$ is the shock radius.  

In the above, $H_0$ is the ordered  
$H_Z$, $H_R$, and $H_\Phi$ that arises from the spherical inflow of the magnetized matter.
The transition to a turbulent magnetic field is assumed to
be rapid on the problem time scale.  For the lowest initial angular velocity that produced an explosion,
$\omega = 0.071$ s$^{-1}$, the integrated growth rate was $I \approx 30$
at a post bounce time of $t_{pb} = 0.14$   s and increased to
$I \approx 90$ by $t_{pb} = 0.25$   s. 

 The initial field was selected so that the final neutron star field will be 
 $\approx 10^{12}$ Gauss.  
The  ordered field
 is thus take to be $H_0 = 10^{12} (10~{\rm km}/r)^2$ Gauss.
As demonstrated in BH98, the field is assumed to grow until near  equipartition,
 \begin{equation}
 \frac{H_{max}^2}{8 \pi \rho} \equiv  \frac{\omega^2 r^2}{4}~~. \end{equation}

Energy is deposited in matter after the field surpasses the $H_{max}$.
Hence, after  $\tilde H = H_{max}$ we let,
\begin{equation}
\dot \epsilon_{matter} = 2 \omega H^2_{max}/8 \pi \rho~~.
\label{epsmag}
\end{equation}
The thermal matter 
pressure $ P_M(\epsilon, \rho)$ above the PNS  is thus 
augmented by Eq. \ref{epsmag} through the increase in $\epsilon_{matter}$.  

In addition, however, there are contributions from
rotation and magnetic effects.
In cylindrical coordinates  the rotational energy density
is just
\begin{equation}
E_{rot} = \frac{1}{2} \rho \omega^2 R^2~~.
\end{equation}
We note that $r = R \sin{(\theta)}$ and  deduce an effective
isotropic pressure due to rotational energy  from an angular
average of the rotational energy,
\begin{equation}
P_{rot} = \frac{ E_{rot}}{3}   = \frac{1}{6} \rho \omega^2 r^2~~.
\end{equation}

Similarly, the energy density due to the isotropic turbulent magnetic
field $\tilde H$ is 
\begin{equation}
\tilde E_{mag} =\frac{\tilde H^2} {8 \pi} ~~,
\end{equation}
The average pressure is then a third of this as is usual for 
an isotropic massless  field.  However, we then reduce this by 
another factor of two due to the solid angle of the directional flow 
of the accreting material.  Hence we write
\begin{equation}
P_{rot} = \frac{\tilde H^2}{48 \pi}~~.
\end{equation}

Calculations were made with different initial angular velocities to 
find out how much rotation was needed to result in an explosion.
These conditions are summarized in Table \ref{table:1}.

\begin{deluxetable}{lr}
\tablenum{1}
\tablewidth{0pt}
\tablecaption{Explosion times and angular velocities}
\tablehead{
\colhead{$\omega_0$ (  s$^{-1}$)}  & $t_{pb}$ {ms} \nl
}
\startdata

$0.20$  & 200 \nl
$0.10$ & 250 \nl
$0.0707$ & 340 \nl
$0.05$ & $\infty$  \nl
 
\enddata
\label{table:1}
\end{deluxetable}

Figure \ref{fig:1} illustrates the effects on a rotating collapse
simulation with and without effects of the magnetic turbulent
pressure contribution.  For this example, 
the rejuvenation of the shock due to
magnetic field amplification above the PNS is clearly demonstrated.

\subsection{Magnetic Bubble Driven Explosion below the PNS Surface}
The nonhomologous collapse of a uniformly rotating
iron core leads to differential rotation and the build
up of a toroidial magnetic field (see Wheeler, Meier \& Wilson 
2002 for a general discussion and references).  The toroidial 
field builds up to  large values and produces
a region unstable to magnetic buoyancy and tension as we now describe.

The magnetic bubble
 model assumes that  the magnetic fields evolve in matter
     which can be treated as having perfect conductivity.
Also,  the magnetic braking back reaction of the field on the
     fluid is not explicitly included, though we estimate its effect. 
We start with the  18 M$_\odot$ stellar progenitor
     model  of Woosley \& Weaver (1985)  at the time at which the iron
     core has just  become unstable to infall. 
While the hydrodynamics is assumed to
     evolve spherically,  the magnetic field is assumed to be axially
     symmetric and is evolved in cylindrical coordinates. 
In what follows, therefore, we will need to simultaneously consider quantities in
 both cylindrical and spherical 
coordinates.  Hence, we use capital $R,~Z,~\Phi$ to distinguish
cylindrical coordinates, while  $r, \theta$, and $\Phi$ are used to denote 
for spherical coordinates. 
 We assign all the matter
     with an angular rotational velocity $\omega_0$  about the $Z$ axis. A uniform
     magnetic field in the $Z$ direction is assumed. Each mass shell is then zoned
     in the $\theta$ direction for the calculation of the magnetic fields and
rotational motion.

  The magnetic flux in the $Z$ direction is taken as constant
in $\Phi$  for each ($r,\theta$)
     zone. Each mass shell rotates rigidly and preserves its angular momentum.
     As described in Section 3, 
these assumptions lead to the following equation for the evolution of the
     the toroidal field.
\begin{eqnarray}
\dot H_\Phi  &=& H_Z^0 \omega_0 \biggl(\frac{r_0}{r}\biggr)^4 
\sin{(\theta)} \cos{(\theta)} 
\biggl( \frac{dr_0}{dr} \frac{r}{r_0} - 1\biggr) \nonumber \\ 
&&- \frac{H_\Phi}{r} \frac{\partial (r v)}{\partial r}
~~.
\label{hphi}
\end{eqnarray}
Note, that the $\sin{(\theta)}\cos{(\theta)}$ product implies a maximum
rate of field growth  in a region inclined $45^o$ from the rotation axis.

      The poloidal field components then follow from flux conservation,
\begin{equation}
\dot H_Z  = H_Z^0  \biggl(\frac{r_0}{r}\biggr)^2 
\biggl(\sin^2{(\theta)} 
 \frac{dr_0}{dr} \frac{r}{r_0} + \cos^2{(\theta)} \biggr) 
~~.
\label{hz}
\end{equation}
\begin{equation}
\dot H_R  = H_Z^0  \biggl(\frac{r_0}{r}\biggr)^2 
\sin{(\theta)}\cos{(\theta)} 
\biggl( \frac{dr_0}{dr} \frac{r}{r_0} - 1\biggr) 
~~,
\label{hr}
\end{equation}
      where $r_0$  is the initial radius of a mass shell and $r$ is the shell radius
      at a later time. The quantity, $H_Z^0$,  is the initial uniform magnetic field and $\omega_0$  
      is the initial angular velocity.

The initial angular rotational velocity
     was chosen to be large enough that a toroidal field will build up
     to a size such that the buoyancy will be large enough to over come
     the stabilizing outward increasing entropy gradient of the matter.  Such
 buoyancy will then
     cause matter to turnover by the quasi-Ledoux convection
(Wilson \& Mayle 1993; Wilson \& Mathews 2003).
This turnover then  brings $\nu$-rich material
     to the surface.  This enhances the neutrino luminosity enough 
at early times to achieve a successful  explosion.

\section{Results}

          Here, we present results for a plausible model in which
the initial precollapse magnetic field, $H_Z$, was chosen to be $3.16 \times 10^7$ G.
        This field magnitude was chosen because it was estimated that the resulting
        neutron star would have a magnetic field of the order of $10^{12}$ G.
        Several initial angular velocities were tried in order to find the minimum
        amount of rotation required to produce an explosion. The minimum angular
        velocity was found to be 0.3 rad   s$^{-1}$ for this magnetic field strength.

This angular velocity is rather high.  Nevertheless preliminary 
axially-symmetric hydrodynamics calculations 
(Tipton 2004) of the collapse of
a star with $\omega = 0.3$ s$^{-1}$ have found only a small distortion
from sphericity for the first 0.2 s after bounce, though a vortex would form unless
viscosity was included.
 
As noted above, this angular velocity leads to a
        rotation rate than is comparable to the shortest
observed  period pulsars and is near the maximum
Keplerian limit on neutron-star spin and is somewhat larger than the rotation
rate observed in young pulsars (Manchester 2004).  
Nevertheless, massive stars are known 
(e.g. Penny, Sprague \& Seago 2004)
to have high surface  rotation velocities, typically $\sim 200 $ km s$^{-1}$.
If these stars rotate uniformly, then unless significant angular momentum transfer occurs
during collapse, they would form a neutron star near the maximum rotation rate.

         We also point out, that  lower spins are required 
if a higher  initial magnetic field is adopted.  For example,
a calculation was made with a smaller  initial  rotation rate
        of 0.1 rad   s$^{-1}$ and a higher precollapse  magnetic field of $10^8$ G. 
The final magnetic field was correspondingly higher ($3\times 10^{12}$ G).
A good explosion resulted and the final neutron star period increased to a few ms.
With this scaling a final field of $10^{13}$ G
only requires an initial  rotation rate of 0.03 s$^{-1}$.
Field strengths of $\sim 10^{13}$  are comparable to that observed in
 a large number of pulsars (Manchester 2004), e.g the Crab pulsar for which 
$H \approx 8\times 10^{12}$ Gauss.  Indeed, it is by now well established that
magnetars with fields as high as $10^{15}$ G exists.  Such stars
would require only rather small rotation rates to induce a supernova.

        As the inner core of the star collapses in our bench-mark model,
 the rotation and magnetic fields cause the  collapse to become very non-homologous. 
Density in the inner region quickly rises from  $\rho \approx \sim 4 \times 10^9$ to
      a bounce density of $\sim 5 \times 10^{14}$  g cm$^{-3}$. Above a baryonic mass cut of 
about 1.5  M$_\odot$
      the density rises very slowly. This leads to large values of $r_0/r$
      and $(dr_0/dr)(r/r_0)$. Hence, a large $H_\Phi$ field is developed
according to Eq. (\ref{hphi}). The componendt $H_R$ and
      $H_Z$ rise as well but not nearly as much as $H_\Phi$. The azimuthal
$H_\Phi$ component thus  becomes the dominant
      field component in the proto-neutron star. 

\begin{figure}
\psfig{file=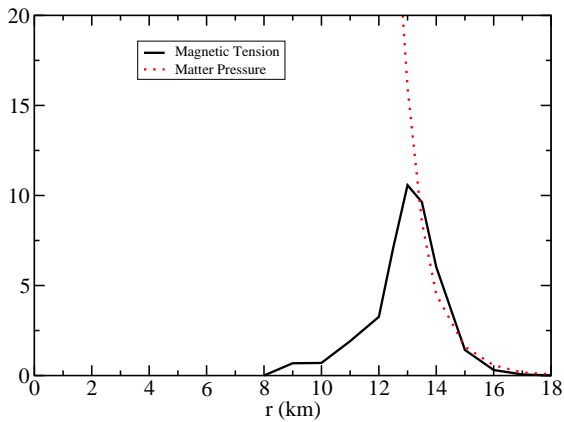,width=3.5in,angle=270}
\vskip 1. in
\caption[]{ Magnetic tension (in units of $10^{30}$ erg cm$^{-3}$)
and matter pressure (in  units of $10^{31}$ erg cm$^{-3}$)  vs radius along 
a line inclined at 45$^o$ to the rotation axis.  Note that the magnetic field is
concentrated near the neutrino-sphere which for this model is at 16.5 km.
This figure represents the case of an initial
magnetic field of $3.16 \times 10^7$ Gauss and a rotation rate of
$0.3$ rad s$^{-1}$ at a time 0.6   s after core bounce.
}
\label{fig:2}

\end{figure}

A key result from these simulations is that the principle field energy density, 
$H_\Phi^2/8 \pi$, forms two concentrated toroidal shaped regions at about 45$^o$ from the rotation 
axis.  Figure \ref{fig:2} compares the magnetic tension $H^2/4 \pi$ with the
matter pressure along a 45$^o$ radius. 
The  buoyancy of and tension of these
      toroidal regions will  stir the matter and induce the transport of matter and  neutrinos
from the core to the surface.  We here endeavor to provide an estimate of the
effects of this transport on  the explosion.

The motion    of material due to these combined effects of rotation, magnetic buoyancy
and magnetic tension is exceedingly complex and would require a fully three dimensional
MHD code.  Nevertheless, the essential features of this magnetic convection
can be deduced via a
a diffusion algorithm  which obeys all relevant conservation laws as it transports
matter, radiation, and neutrino properties.
The same algorithm as has been employed for neutron-finger 
convection in the  Livermore supernova model (Wilson \& Mathews 2003).
This allows an easy  comparison with results from that mechanism.
 
The effect of the magnetic buoyancy and tension instability is therefore  modeled as follows.
When the buoyancy of the magnetic field is sufficient to overcome the positive entropy gradient
then a magnetic diffusion algorithm  is initiated. 

The problem of how  the magnetic bubbles will rise and exchange energy, etc., with
its surroundings is quite difficult to solve directly.  To model the effect
we use the existing convection algorithm in the
supernova models to transport energy, composition, and neutrinos.
Dimensional analysis is used to set the size of an effective diffusion coefficient,
i.e.~the effective diffusion coefficient
is taken as scaling with the product of a mixing length times
the characteristic magnetic velocity parameter.
\begin{equation}
D = \frac{l}{30} \sqrt{\frac{H^2}{8 \pi \rho}}~~,
\end{equation}
       where $H$ is the maximum of the magnitude of the $H_\Phi$ field and
\begin{eqnarray}
l &=& r_{\nu} - r ~~,~~ r < r_H \nonumber \\
 &=& r_{\nu} - r_H  ~,~ r \ge  r_H ~,
\end{eqnarray}
       where $r_H$ is the radius at which the maximum of $H$ occurs, and $r_\nu$ is the radius
       of the neutrino photosphere. The factor of 30 in the denominator 
was deliberately assumed to be large
       so as to be conservative in our estimate of the size
and behavior of the magnetic convective
cells.  A smaller cell size implies a slower convective lifetime.
It should be noted that our calculation is only good enough to show
the scale of rotation and magnetic field that can initiate an explosion.

As in the case of the neutron-finger
instability (Wilson \& Mathews 2003), 
the magnetic  convection brings up proton-rich matter (compared to the
deleptonized surface regions)  as
       well as neutrinos towards the surface of the proto-neutron star and results
       in an enhanced neutrino luminosity soon after bounce.

\begin{figure}
\psfig{file=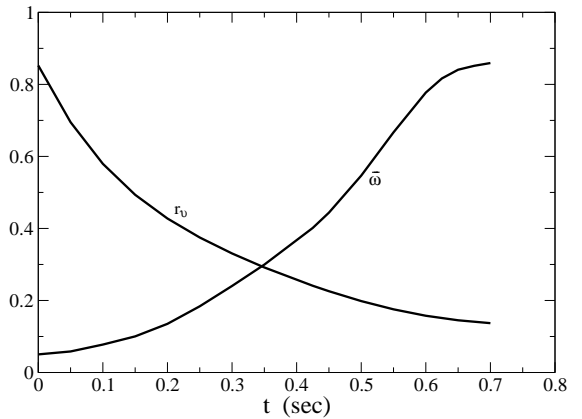,width=3.5in,angle=270}
\caption[]{ Curve marked $r_\nu$ shows  the 
 neutrino photospheric radius in units of 10$^2$ km versus time
as the neutron star relaxes to a radius of $\sim 10$ km.
The curve labeled $\bar \omega$ is the 
    angular speed  (in units of $10^4$ rad   s$^{-1}$)
 of the     proto-neutron star averaged for matter inside 
$r_\nu$  as a function of time.  }
\label{fig:3}

\end{figure}

\begin{figure}
\psfig{file=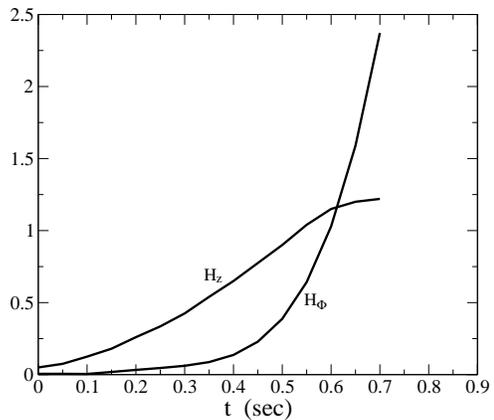,width=3.5in,angle=270}
\caption[]{
Curve $H_Z$ shows the maximum of the $Z$-component of the magnetic field 
as a function of time in unite of $10^{12}$ G.
Curve  $H_\Phi$ is the maximum value of the $\Phi$-component of the magnetic field
as a function of time in units of $10^{16}$ G.
}
\label{fig:4}
\end{figure}

\begin{figure}
\psfig{file=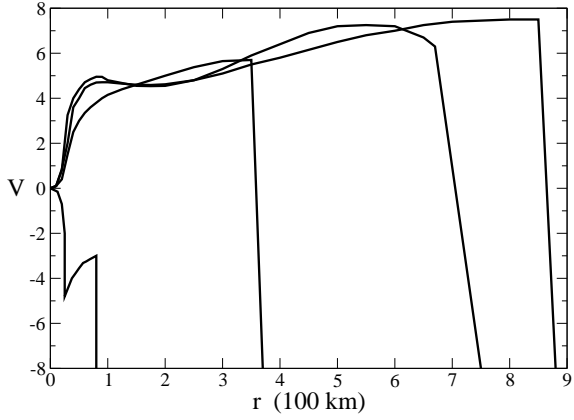,width=3.5in,angle=270}

\caption[]{

Radial velocity in units of $10^3$ km   s$^{-1}$ at various indicated times
post bounce (from left to right) 
of 0.60, 0.63, 0.66, and 0.68   s for the model with magnetic convection. 
}
\label{fig:5}
\end{figure}

\begin{figure}
\psfig{file=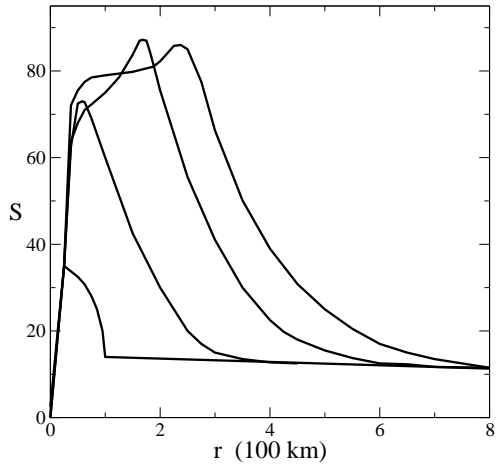,width=3.5in,angle=270}
\caption[]{
Entropy per baryon $S$ versus radius for the post-bounce times
(from left to right) 
of 0.60, 0.63, 0.66, and 0.68   s for the model with magnetic convection.
}
\label{fig:6}
\end{figure}

\begin{figure}
\psfig{file=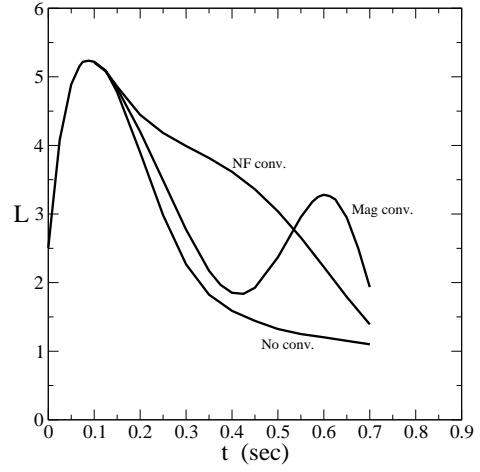,width=3.5in,angle=270}
\caption[]{
Electron neutrino luminosities in units of $10^{52}$ erg   s$^{-1}$
as  a function of post-bounce  time for three calculations as labeled:
no convection; neutron-finger convection, and magnetic convection.
Note, that the early short-duration shock break-out luminosity has been
suppressed.
}
\label{fig:7}
\end{figure}

\begin{figure}
\psfig{file=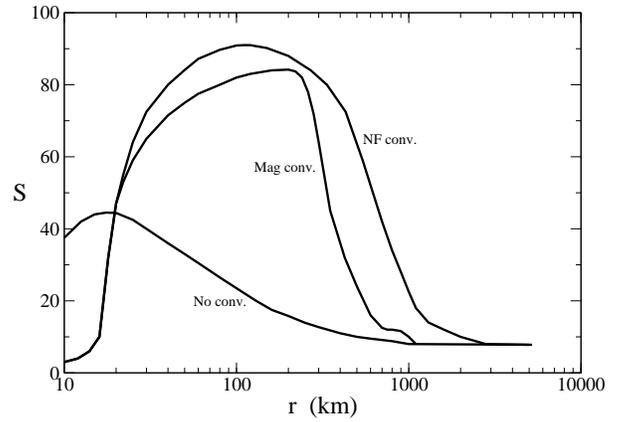,width=3.5in,angle=270}
\caption[]{
Entropy  per baryon versus radius at a post-bounce time of 0.68 s for the
cases of no convection; neutron-finger convection, and magnetic convection.
}
\label{fig:8}
\end{figure}

\begin{figure}
\psfig{file=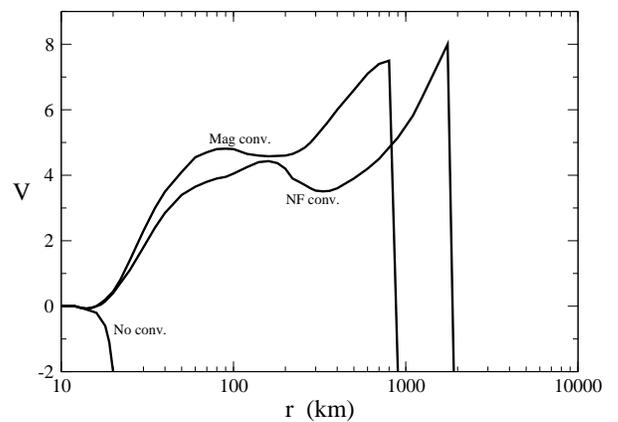,width=3.5in,angle=270}
\caption[]{Velocity in units of $10^3$ km   s$^{-1}$ at a post-bounce time of
0.68   s for the cases of 
no convection; neutron-finger convection, and magnetic convection.
}
\label{fig:9}
\end{figure}

Figures \ref{fig:3}-\ref{fig:6} show some of the 
          details of the calculation with an initial field of $3.16 \times 10^7$ G 
        and a rotation rate of 0.3 rad   s$^{-1}$.  In Figures \ref{fig:7}-\ref{fig:9}
 these are  compared to
        calculations with no convection  and also those with neutron finger convection.
        As can be seen in Figures \ref{fig:7}-\ref{fig:9},
 all three calculations give nearly the same behavior 
until a few tenths of seconds after the core bounce.

In Figure \ref{fig:3} the neutrino photospheric radius $r_\nu$  and 
        the proto-neutron star average angular velocity $\bar \omega $ are presented. 
          The final angular speed is very high, $\omega \approx 9\times 10^3$ rad   s$^{-1}$,
corresponding to a rotation period of $P \approx 1$ ms.  The star, however
will quickly spin down.  The surface
        field  $H_\Phi$ is very high and it is anchored in the massive non-rotating
        envelope. The magnetic torque should be of order,
\begin{equation}
\tau_H \approx \int \biggl( \frac{\partial H_\Phi}{\partial R} H_Z - 
\frac{\partial H_\Phi}{\partial Z} H_R \biggr ) R r^2 dr~~.
\end{equation}
Putting in numbers for $ H_\Phi, H_R,$ and $H_Z,$ from the simulations,
we obtain $\tau_H \approx$ a few $\times 10^{46}$ erg.
Then for a rotational moment of inertia of the nascent neutron star of
$I \approx 10^{45}$ g cm$^{2}$ one has,
\begin{equation}
\frac{\dot \omega}{\omega} = \frac{\tau_H}{I \omega} 
\approx ~{\rm a~few}~ \times 10^{-3}~~{\rm s}^{-1}~~.
\end{equation}
Hence the magnetic torque should be
able to slow the spin of the neutron star
        considerably within a several minutes.
The maximum values of the
        magnetic fields $H_Z$, $H_\Phi$  as a function of time are shown in Fig.~\ref{fig:4}.
 This figure shows that it
only takes several tenths of seconds to get large magnetic  fields. 

Figures  \ref{fig:5} and \ref{fig:6},  show the radial velocity
        and entropy per baryon, respectively, at various times  for the mass shell
with the highest outward velocity in  a model with magnetic convection and
no neutron-finger instability. 
Here it is apparent  from the expanding radii and increasing entropies
that a good explosion has resulted.
        When the entropy rises to about 80 the heating by neutrino-electron scattering is
        equal to that of neutrino capture. Hence,  even though the luminosities are
        falling the heating within the high-entropy bubble will remain substantial.

 Figure \ref{fig:7} compares the electron
        neutrino luminosities as functions of time for three runs: no convection,
neutron-finger convection, and magnetic-field driven convection. 
Here we see that in the case of the neutron-finger convection, luminosity comes
        on early but is eventually surpassed by the magnetic luminosity. The
        entropy profiles in radius at a time of 0.68   s post-bounce seconds are shown
        in Figure \ref{fig:8}. The entropy in the magnetic case is only slightly less than of
        the neutron-finger case. From Fig.~\ref{fig:9} we see that the outward velocity  for the
        magnetic calculation is only slightly below  the neutron-finger velocity.

\section{Conclusions}

Although the  Livermore supernova model
with neutron-finger convection is a viable description of core-collapse supernovae,
the present calculations suggest an alternative to the neutron-finger instability
for initiating an explosion.
If correct, this may provide the long sought after insight as to how core-collapse
supernovae become sufficiently heated behind the shock to explode. 
 An interesting possible side effect of the magnetic field generation that we studied is
         that an axial jet and/or  a prolate bulge in the  mass distribution 
should arise independently of how the convection is driven.
         Such features, for example,  might be an explanation of
the observation that most remnants emit polarized optical radiation.

 It should be noted that both this magnetic turbulence effect and the
magnetic bubble formation below the proto-neutron star surface will act together 
with neutron-fingers to induce an explosion.
Clearly, more detailed work utilizing 
two and three dimensional MHD simulations is required
to explore whether the magnetic buoyancy effect described herein is sufficient to
induce an explosion.  
This  is, however, a difficult and time-consuming calculation.
It is hoped the present work will stimulate further effort to understand this
possibly important contribution to the complex paradigm of core-collapse supernovae.

\acknowledgments 
Work at the Lawrence Livermore National Laboratory performed in part
under the auspices of the U.~S.~Department of Energy under contract
W-7405-ENG-48 and NSF grant PHY-9401636.
 Work at the
University of Notre Dame supported by the U.S. Department of Energy
under Nuclear Theory Grant DE-FG02-95-ER40934. We acknowledge 
John Hawley  for suggesting that we look at the effects of magnetic turbulence 
as a means to initiate a supernova explosion.
The authors wish to thank R. Tipton of LLNL for useful discussions.
They also wish to acknowledge Jonathan Wilt for help in the preparation
of part of this manuscript who is also supported in part through
an NSF Research Experience for Undergraduates grant at the Univ. of Notre Dame.
One of the authors (JRW) also acknowledges many useful related discussions
which took place at the Aspen Center for Physics.


\begin{thebibliography}{}
\bibitem{Akiyama} Akiyama, S.,  Wheeler, J.C.,  Meier, D.L. \& Lichtenstadt, I. 
2003, ApJ, 584,  954.

\bibitem{Hawley88} Balbus, S. A. \& Hawley, J. F. 1988, Rev. Mod. Phys., 70, 1

\bibitem{Bruenn85} Bruenn, S.W.  1985, ApJ Suppl. {\bf 58},  771


\bibitem[{Bruenn(1993)}]{bruenn93}
Bruenn, S.~W. 1993, in Nuclear Physics in the Universe, ed. M.~W. Guidry \&
  M.~R. Strayer, Proceedings of the First Symposium on Nuclear Physics in the
  Universe held in Oak Ridge, Tennessee, USA 24-26 September 1992 (Bristol: IOP
  Publishing), 31

\bibitem{bruenn04}
Bruenn, S.~W., Raly, A. \& Mezzacappa, A. 2004, ApJ, submitted, astro-ph/0404099

\bibitem{buras}
Buras, R., Rampp, M., Janka, H.~T., \& Kifonidis, K. 2003, PRL, 90, 241101 

\bibitem{burgio} Burgio, G. F.,  Schulze, H.-J. \&  Weber, F.  2003, A\&A 408, 675

bibitem{burrows} Burrows, A. 2004,
in {\it Proceedings of the Twelfth Workshop on "Nuclear Astrophysics,
a Tribute to an Explosive Astrophysicist, Wolfgang
Hillebrandt, on the occasion of his 60th Birthday}, Ringberg Castle, L
ake Tegernsee,
Germany, March 22 - 27, 2004, eds. E.  Muller and H.-Th. Janka
on the Physics of the r-Process, Seattle (2004)
(World Scientific: Singapore) in press, astro-ph/0405427.

\bibitem{burrows95} Burrows, A., Hayes, J. A. \& Fryxell, B. A. 1995, ApJ, 450, 830

\bibitem{cc} Cardall, C. Y. 2004, in {\it The r-Process: The Astrophysical Origin of the
Heavy Elements and Related Rare Isotope Accelerator
Physics}, Y. z. Qian, et al. , Eds., 
(World Scientific: Singapore) pp. 186-195

\bibitem{Fryer} Fryer,  C.L. \& Heger, A. 2000, ApJ,   541,  1033


%

\bibitem[{Liebend\"{o}rfer {et~al.}(2001)Liebend\"{o}rfer, Mezzacappa, Messer,
  Hix, Thielemann, \& Bruenn}]{liebendorfermtmhb01}
Liebend\"{o}rfer, M., Mezzacappa, A., Messer, O.~E.~B., Hix, R.~M., Thielemann,
  F.-T., \& Bruenn, S.~W. 2001, PRD, 63, 103004

\bibitem{Manchester} Manchester, R. S. 2004, Science, 304 542

\bibitem{Mezzacappa01} Mezzacappa, A., Liebend\"{o}rfer, M., Messer, O.~E.~B., 
Hix, R.~M., Thielemann,
  F.-T., \& Bruenn, S.~W. 2001, Physical Review Letters, 86, 1935


\bibitem{Penny}
Penny, L.~R., Sprague, A.~J.~\& Seago, G.~ 2004, ApJ, 617, 1316

\bibitem{Phinney}
Phinney, E. S. \& Kulkarni, S. R. 1994, ARA\&A, 32, 591



\bibitem{rampp00}
M. Rampp, M.,  \& Janka, H.-Th. 2002, ApJL, 539, L33   

\bibitem{rampp02}
M. Rampp, M.,  \& Janka, H.-Th. 2002, A\&A, 396, 361


\bibitem{Thompson03}Thompson, T.A.,  Burrows, A. \& Pinto, P.A. 2003, ApJ,   592,  434.

 
\bibitem{Tipton} Tipton, R. 2004, Priv. Comm.


\bibitem{weber99}Weber, F. 1999, J. Phys. G,  25, 195

\bibitem{Wheeler}Wheeler, J. C., Meier, D. L. \&  Wilson, J. R. 2002, ApJ, 568, 807

\bibitem{WM03} Wilson, J.~R.~\& Mathews, G.~J., 2003, {\it Relativistic Numerical
Hydrodynamics}, (Cambridge University Press, Cambridge).

\bibitem{WM88} Wilson, J.~R.~\& Mayle, R. W. 1988, Phys. Rep., 163, 63.

\bibitem{WM93} Wilson, J.~R.~\& Mayle, R. W. 1993, Phys. Rep., 227, 97.





\bibitem[Woosley \& Weaver(1995)]{1995ApJS..101..181W} Woosley, S.~E.~\& 
Weaver, T.~A.\ 1995, \apjs, 101, 181 





\bibitem{yamada99} 
Yamada, S., Janka, H.-Th., and Suzuki, H. 1999, A\&A, 344, 533

%

\end{thebibliography}
\end{document}